\documentclass[aps,prl,reprint,showpacs,superscriptaddress,longbibliography]{revtex4-2}
\usepackage{mathtools,amssymb,graphicx,units}

\usepackage[plainpages=false,pdfpagelabels,colorlinks=true,linkcolor=red,urlcolor=blue,citecolor=blue,pdftitle={Title},pdfauthor={Rubem Mondaini},pdfdisplaydoctitle=true,pdfduplex=DuplexFlipLongEdge]{hyperref}
\usepackage{amsmath,graphics,units}
\usepackage[table,xcdraw]{xcolor}
\usepackage{verbatim}
\usepackage[english]{babel}
\usepackage{color}
\usepackage{ulem}
\usepackage{physics}

\newcommand{\tcr}{\textcolor{red}}

\newcommand{\beginsupplement}{%
        \setcounter{table}{0}
        \renewcommand{\thetable}{S\arabic{table}}%
        \setcounter{figure}{0}
        \renewcommand{\thefigure}{S\arabic{figure}}%
}

\begin{document}
\title{Two-dimensional polarized superfluids under the prism of the fermion sign problem}

\author{Tian-Cheng Yi}

\affiliation{Key Laboratory of Optical Field Manipulation of Zhejiang Province, Physics Department, Zhejiang Sci-Tech University, Hangzhou 310018, China}
\affiliation{Beijing Computational Science Research Center, Beijing 100193, China}
\author{Song Cheng}
\affiliation{Beijing Computational Science Research Center, Beijing 100193, China}
\author{Ian Pil\'{e}}
\affiliation{HSE University, 101000 Moscow, Russia}
\author{Evgeni Burovski}
\affiliation{HSE University, 101000 Moscow, Russia}
\author{Rubem Mondaini}
\email{rmondaini@csrc.ac.cn}
\affiliation{Beijing Computational Science Research Center, Beijing 100193, China}

\begin{abstract}
Understanding if attractive fermions in an unbalanced occupation of its flavors can give rise to a superfluid state in two dimensions (2D), realizing the Fulde-Ferrel-Larkin-Ovchinnikov (FFLO) state, presents a long-standing question. A limitation on its solution by numerics is posed by the sign problem, which constrains the applicability of quantum Monte Carlo techniques at sufficiently low temperatures and large lattice sizes, where a potential signature of polarized superfluidity would be unambiguous. By using a recently explored argument that the sign problem may be used instead to infer quantum critical behavior, we explore the regime where partial polarization occurs in the phase diagram, further showing that the average sign $\langle {\cal S}\rangle$ of quantum Monte Carlo weights tracks the criticality between balanced (or fully polarized) and polarized phases. Using the attractive Hubbard model with an unbalanced population, our investigation expands the scope of problems in which $\langle {\cal S}\rangle$ can be used for monitoring critical behavior, providing compelling albeit indirect evidence for the robustness of an FFLO phase in 2D. 
\end{abstract}

\maketitle
\paragraph{Introduction.---}The Bardeen-Cooper-Schrieffer mechanism~\cite{Bardeen1957} is a triumph of theoretical physics. It describes how fermionic pairs condense, inducing superconductivity via a degenerate occupancy of zero momentum pairing modes, and ultimately explains the observed physical properties of various superconducting materials, including quantitative evidence for critical temperatures and the Meissner effect. Right after its introduction, Fulde and Ferrel~\cite{Fulde1964}, and independently Larkin and Ovchinnikov~\cite{Larkin1964}, proposed schemes at which pairing can still occur, but at finite-momentum instead, to accommodate an imbalance of fermionic species, i.e., a finite polarization wherein superconductivity and net magnetism can coexist.

Evidence suggestive of this latter scenario in condensed matter settings now prevails, including in heavy fermions compounds~\cite{Bianchi2003, Radovan2003, Koutroulakis2010} and organic superconductors~\cite{Uji2006, Lortz2007, Beyer2012, Koutroulakis2016, Wosnitza2018, Imajo2021}. Yet, the most compelling observations emerge in the controlled environment of trapped ultracold atoms where different electronic spin flavors are mapped into fermionic atoms with different hyperfine states~\cite{Zwierlein2006, Partridge2006, Partridge2006b, Zwierlein2006b, Shin2006, Shin2008, Liao2010}. In the one-dimensional case~\cite{Liao2010}, it confirms the Bethe ansatz results under local density approximation in the presence of a trap~\cite{Orso2007, Hu2007}: A phase separation of a partially polarized (PP) core with a fully paired wing emerges, but no direct evidence of superfluidity nor microscale phase separation associated with the two-momentum components in the superfluid order parameter of the PP regime~\cite{Kinnunen2018} have been so far experimentally identified~\cite{Bloch2010}.

Nonetheless, unbiased methods such as the density matrix renormalization group (DMRG)~\cite{Feiguin2007, Tezuka2008, Luscher2008, Rizzi2008} or quantum Monte Carlo (QMC) calculations~\cite{Batrouni2008} applied to lattice models unambiguously describe the zero-temperature ($T=0$) FFLO superfluid and its ensuing quasi-long range order~\cite{Frahm1990, Frahm1991}. Extensions to finite-$T$'s corroborate the real-space modulated pairing correlations, characteristic of this exotic state, in an extensive regime of parameters~\cite{Wolak2010}.

Much less is known, however, about the actual fate of the ground-state FFLO phases in two dimensions, apart from mean-field solutions~\cite{Parish2007}. Controlled analytical methods are not available, DMRG simulations are inherently constrained to quasi-one-dimensional geometries such as ladders or cylinders~\cite{Feiguin2009, Potapova2023}, and QMC calculations often face the exponential barrier of the sign problem (SP)~\cite{Loh1990, Troyer2005}. Bypassing it, Ref.~\cite{Vitali2022} established that a $T=0$ FFLO state emerges in QMC calculations in two dimensions by setting a gauge condition constraining the path of the stochastic integration~\cite{Zhang1997}. At $T\neq 0$ without constraint, despite observing pairing indication with finite-momentum~\cite{Wolak2012}, no direct manifestation of phase coherence (and thus superfluid behavior) is attainable in this type of simulation precisely because of the emergence of the SP. Lastly, by employing a diagrammatic QMC method, which replaces the SP challenge by a truncation on the order of Feynmann diagrams being sampled~\cite{Kozik2010}, recent results have shown the FFLO instability in a small regime of parameters (quarter-filling and small polarization)~\cite{Gukelberger2016}.

While the SP poses a fundamental obstacle in extracting the physical information needed to understand the putative formation of a polarized superfluid and its ensuing modulation of the order parameter in real space~\cite{Kinnunen2018}, recent studies have suggested that critical properties can be directly imprinted in it~\cite{Mondaini2022, Mondaini2022-2}. Our goal is to show that FFLO physics precisely falls into this category, and the informed analysis of the properties of the weights in the Monte Carlo sampling can aid in establishing a comprehensive phase diagram of the relevant lattice model where polarized superfluidity can occur. 

\paragraph{Model and argument.---} The minimal model to describe such physics in a lattice is the fermion-Hubbard model,
\begin{align}
    \hat {\cal H} = -t\sum_{\langle i,j\rangle,\sigma}\hat c_{i\sigma}^\dag \hat c_{j\sigma}^{\phantom{\dag}} 
    -\sum_{i,\sigma} \mu_\sigma \hat n_{i\sigma} 
    + U\sum_{i}\hat n_{i\uparrow} \hat n_{i\downarrow}\ ,
    \label{eq:H}
\end{align}
where $\hat c_{i\sigma}^\dag(\hat c_{i\sigma}^{\phantom{\dag}})$ is the fermionic creation (annihilation) operator at lattice site $i$ with spin $\sigma=\uparrow,\downarrow$ and $\hat n_{i\sigma}$ is the corresponding number operator. Hopping with an energy scale $t$ is constrained to nearest neighbor sites $\langle i,j\rangle$ in a square lattice with dimensions $L_x\times L_y$, and the onsite interactions are attractive ($U<0$). A spin-dependent chemical potential $\mu_\sigma$ controls the occupancy of each fermionic flavor: $\mu_\uparrow = \mu + h$ and $\mu_\downarrow = \mu - h$; $\mu$ is the global chemical potential and $h$ a Zeeman energy associated to an in-plane magnetic field applied in the lattice such that no additional orbital effects emerge.

\begin{figure}[t!]
\centering
\includegraphics[width=0.99\columnwidth]{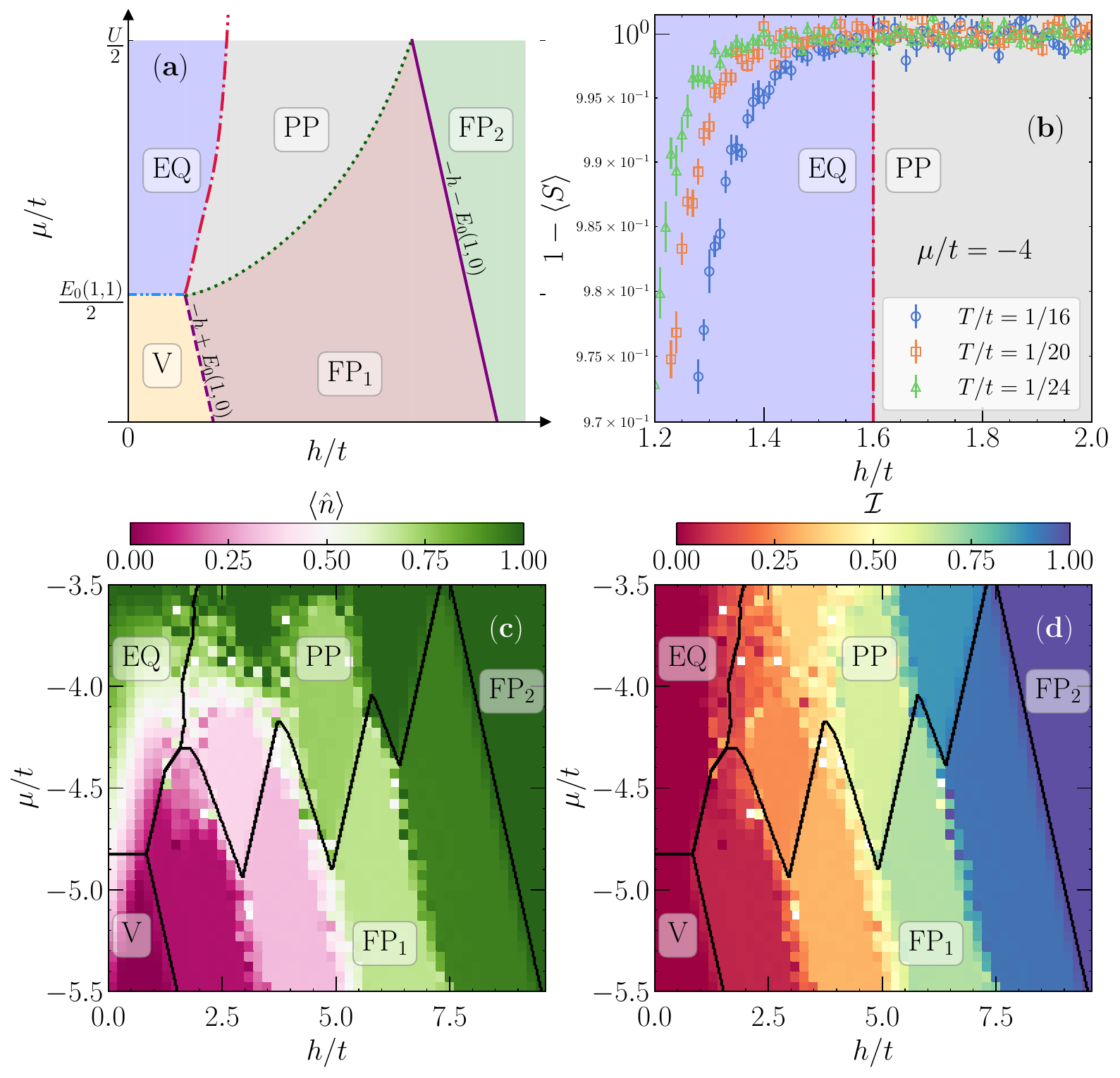}
\caption{(a) Schematic $T=0$ phase diagram of \eqref{eq:H} in the $\mu$--$h$ parameter space; analytical values annotated for the transitions to the vacuum or (saturated) fully polarized phases are explained in \cite{SM}. Panels (b)--(d) show an exact diagonalization vs.~QMC comparison in a $4\times 4$ with periodic boundary conditions (PBC) and $U/t=-7$; exact phase boundaries (lines) are obtained by an energetics analysis of the different types of states~\cite{SM}. (b) The average total sign $\langle{\cal S}\rangle$ at $\mu=-4t$ and different $T$'s. Color maps of the QMC results for (c), the total density $\langle \hat n\rangle$, and (d), imbalance $\cal I$, at $T/t=1/20$ with imaginary-time discretization $t\Delta\tau=0.1$}
\label{fig:fig1}
\end{figure}

Inspired by either mean-field~\cite{Parish2007} or thermodynamic Bethe ansatz~\cite{Cheng2018} phase diagrams, the expected regimes in the $\mu$--$h$ plane are: (i) vacuum (V), where the total electronic density $\langle \hat n\rangle$ is zero [$\langle \hat n\rangle \equiv \sum_{\sigma}\langle \hat n_\sigma\rangle = \sum_{i,\sigma}\langle\hat n_{i\sigma}\rangle/(L_x L_y)]$ with zero imbalance ${\cal I}$ [${\cal I} \equiv \langle \hat n_{\uparrow} \rangle -  \langle \hat n_{\downarrow} \rangle$], (ii) equal (and finite) densities (EQ) with ${\cal I}=0$, (iii) PP where ${\cal I}\neq 0$ and $\langle \hat n\rangle > 0$, and two fully polarized (FP) phases, i.e., wherein the density of the minority flavor is zero, differing on whether the density of the majority one is either (iv) unsaturated, FP$_1$, or (v) saturated, FP$_2$. An exotic polarized superfluid can only thus occur within the PP phase at sufficiently low temperatures [see Fig.~\ref{fig:fig1}(a)].

\renewcommand{\arraystretch}{1.2}
\begin{table}[t!]
\begin{tabular}{c | c | c | c | c}
\hline
 & $\langle {\cal S}\rangle$ &  $\langle {\cal S}_\uparrow\rangle$ & $\langle {\cal S}_\downarrow\rangle$&  ${\cal I}$\\ \hline\hline
 V& 1 & 1 & 1 & 0 \\ \hline
 EQ        & 1 &  $0<\langle {\cal S}_\sigma\rangle< 1$& $0<\langle {\cal S}_\sigma\rangle< 1$ & 0 \\ \hline
 FP$_1$ &  $0<\langle {\cal S}\rangle< 1$ & $0<\langle {\cal S}_\uparrow\rangle< 1$ & 1 & $0<\langle \hat n_\uparrow\rangle<1$ \\ \hline
 FP$_2$& 1 & 1 & 1 & 1 \\ \hline
 PP & $0<\langle {\cal S}\rangle< 1$ &  $0<\langle {\cal S}_\uparrow\rangle< 1$ & $0<\langle {\cal S}_\downarrow\rangle< 1$ & $\neq 0$
\end{tabular}
\caption{Predicted bounds on the average sign of the total and individual weights for each fermionic flavor at sufficiently low temperatures when approaching the thermodynamic limit for each expected phase. The last column gives the corresponding fermionic flavor imbalance.}
\label{tab:ave_sign}
\end{table}

\begin{figure}[b!]
\includegraphics[width=0.99\columnwidth]{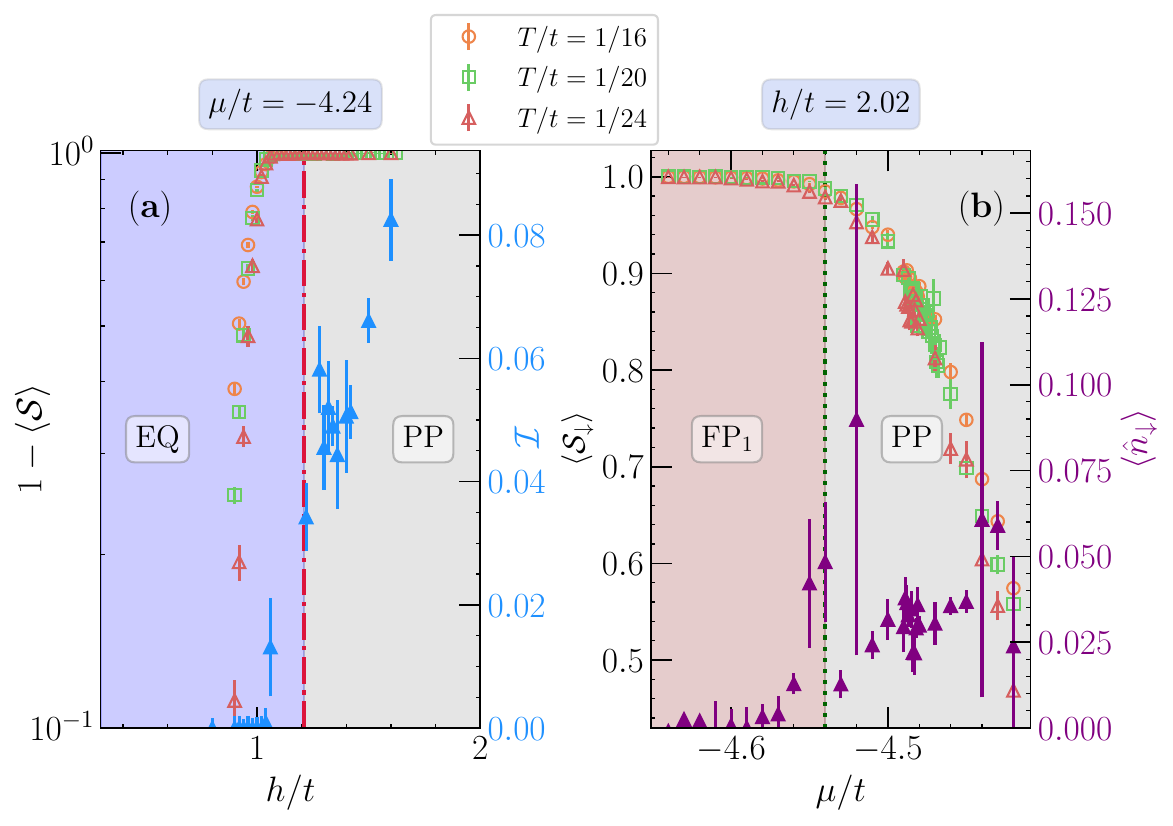}
 \caption{Comparison with DMRG results (vertical lines, see \cite{SM}) in a $40\times 4$ lattice with open boundary conditions. (a) Average sign vs.~$h$ with $\mu/t=-4.24$; also show (right-vertical axis) the corresponding imbalance for $T/t=1/24$. (b) Average sign of the weight of the minority component $\langle {\cal S}_\downarrow\rangle$ 
 vs.~ $\mu$ with $h/t = 2.02$; the right vertical axis gives its average density $\langle \hat n_\downarrow\rangle$, which turns finite in the PP phase.}
 \label{fig:fig2}
\end{figure}

\begin{figure*}[th!]
 \centering
 \includegraphics[width=\textwidth]{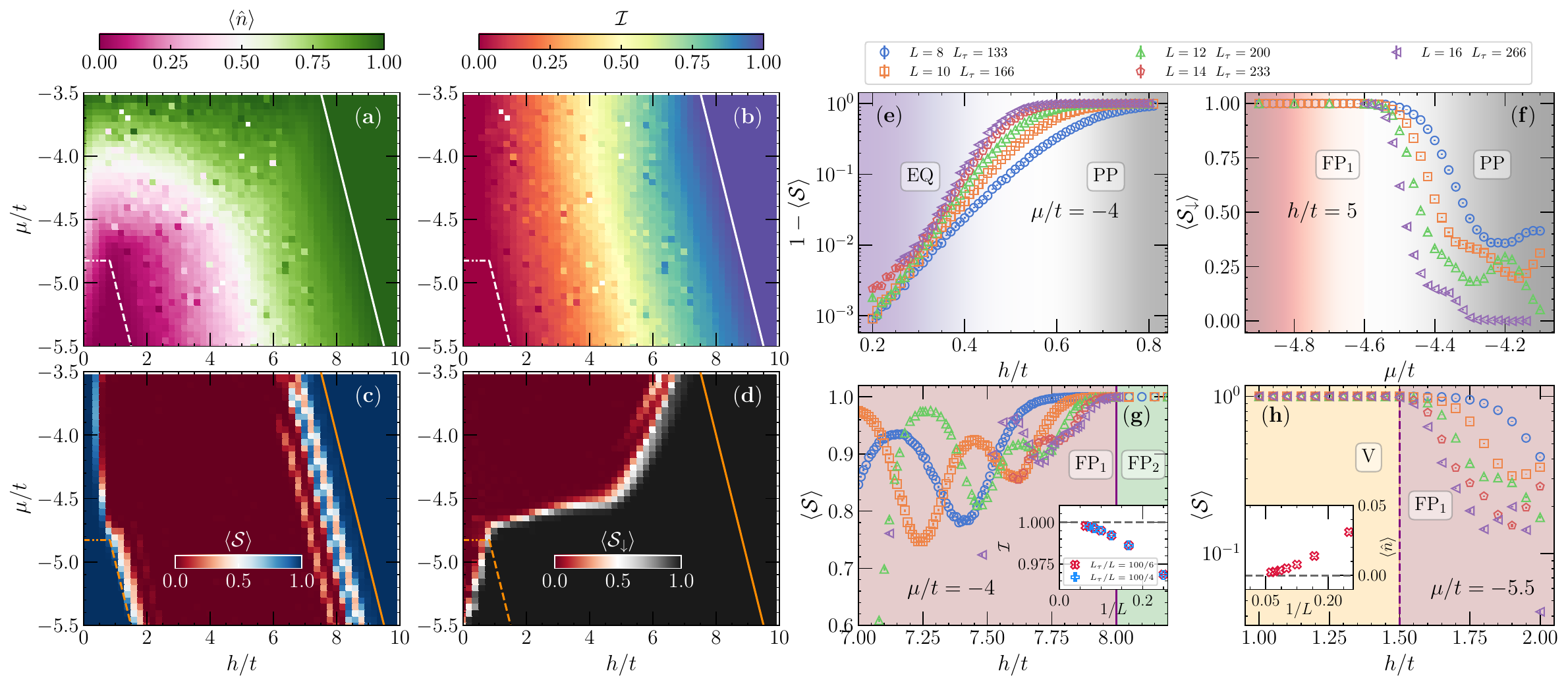}
 \caption{(a) Density $\langle \hat n\rangle$ and (b) imbalance $\cal I$ in the $\mu$--$h$ plane and corresponding total average sign $\langle {\cal S}\rangle$, (c), and the average sign of the fermionic weight for the minority component $\langle {\cal S}_\downarrow\rangle$,  (d), in a $L=16$ lattice with $T/t=1/20$. Lines, as in Fig.~\ref{fig:fig1}(a), mark the boundaries of the trivial phases: vacuum and full polarized FP$_2$. Panels (e--f) depict the phase boundary in details contrasting different system sizes: (e) $1-\langle{\cal S}\rangle$ for $\mu/t=-4$ on the EQ-PP transition, $\langle {\cal S}_\downarrow\rangle$ for $h/t=5$ on the PP-FP$_1$ transition, (g)  $\langle {\cal S}\rangle$ also at $\mu/t=-4$ but large $h$ values on the FP$_1$-FP$_2$ transition, and (h) $\langle {\cal S}\rangle$ vs. $h$ for $\mu/t=-5.5$ marking the vacuum-FP$_1$ transition. Insets in (g) and (h) show the system size extrapolation of $\cal I$ and $\langle \hat n\rangle$ at the analytically obtained transition point to the trivial phases.}
 \label{fig:fig3}
\end{figure*}

Notably, in the auxiliary-field QMC method~\cite{Blankenbecler1981}, the weight $W$ of each configuration in the Monte Carlo sampling can be chosen to be separable in its flavors, $W = W_{\uparrow}(\{x\})\cdot W_{\downarrow}(\{x\})$, where $\{x\}$ is a Hubbard-Stratonovich (HS) field used to decouple the interactions~\cite{Hirsch1983,Hirsch1985}. Knowing that typically the average sign $\langle {\cal S}\rangle \equiv \langle{\rm sgn}(W)\rangle$ decreases exponentially with both the inverse temperature $\beta = 1/T$ and the system size $N=L_xL_y$~\cite{Loh1990,Iglovikov2015}, one can infer that at $T\ll t$, in approaching the thermodynamic limit, the above-described phases can be readily classified according to the values of $\langle {\cal S}\rangle$ or its flavor-dependent counterpart, $\langle {\cal S}_\sigma\rangle$ (see Table~\ref{tab:ave_sign}). For example, phases in which the occupancy $\langle \hat n_\sigma\rangle$ is either 0 or 1 have $\langle {\cal S}_\sigma\rangle = 1$ for whichever field configuration $\{x\}$; this helps in quickly identifying the V and fully polarized phases, FP$_1$ and FP$_2$. On the other hand, since $\langle \hat n_\uparrow\rangle = \langle \hat n_\downarrow\rangle \in (0,1)$ within the EQ regime, one expects that $\langle {\cal S}_\uparrow \rangle =  \langle {\cal S}_\downarrow \rangle \in (0, 1)$ and the total sign $\langle {\cal S}\rangle = 1$~\footnote{We emphasize that we do not have a formal mathematical proof that $\langle {\cal S}\rangle = 1$ in the EQ phase since in principle the determinants for each spin-flavor are not protected to have the same sign with $\mu_\uparrow\neq \mu_\downarrow$ for a given HS configuration, but rather a conjecture based on the exact numerical results, in particular at $4\times 4$ lattice.}. At last, the PP phase is identified by being the only one in which there is no protection, and the average sign of either the total or flavor-resolved weights is $0<\langle {\cal S}\rangle,\langle {\cal S}_\sigma\rangle<1$.
 
\paragraph{Benchmarking.---} To quantify these assertions, we contrast the QMC results against numerically exact ones in a small $4\times 4$ lattice in Fig.~\ref{fig:fig1}. Hereafter, we fix the interaction strength to $U/t=-7$ and set the temperature discretization $T = t/(L_\tau\Delta\tau)$. The phase boundaries extracted from exact diagonalization (ED) can be quickly inferred from the energetics of the different phases in the $\mu$--$h$ plane~\cite{SM} and are quantitatively represented in Figs.~\ref{fig:fig1}(c,d); in such a small lattice, finite-size effects are prominent, leading to less smooth transition curves. Still, both the density [Fig.~\ref{fig:fig1}(c)] and imbalance [Fig.~\ref{fig:fig1}(d)] obtained from QMC at $T/t=1/20$ closely match the exact results at $T=0$. According to Table~\ref{tab:ave_sign}, the transition to a partially polarized regime from the EQ phase can be tracked by the departure of $\langle {\cal S}\rangle$ from 1 at a critical value of the field $h$; that is indeed observed in Fig.~\ref{fig:fig1}(b) at $\mu=-4t$ with consistently low temperatures.
 
A second benchmark can be performed with much larger system sizes, contrasting the QMC results with DMRG in $4\times 40$ ladders. We now focus on two transitions, EQ-PP at fixed $\mu/t = -4.24$ [Fig.~\ref{fig:fig2}(a)] and PP-FP$_1$ at fixed $h/t=2.02$ [Fig.~\ref{fig:fig2}(b)]. As before, the former transition is tracked at low-$T$'s when $\langle {\cal S}\rangle\to 0$ when entering the PP phase. The transition from a partial to fully polarized state is tracked by the average sign of the weight of the minority component $\langle {\cal S}_\downarrow\rangle \to 1$, since $\langle \hat n_\downarrow\rangle\to 0$ when entering FP$_1$. Indeed, such predictions hold fairly well for the QMC results at small temperatures (see \cite{SM} for details).
 
\begin{figure}[htp]\centering
\includegraphics[width=0.5\textwidth]{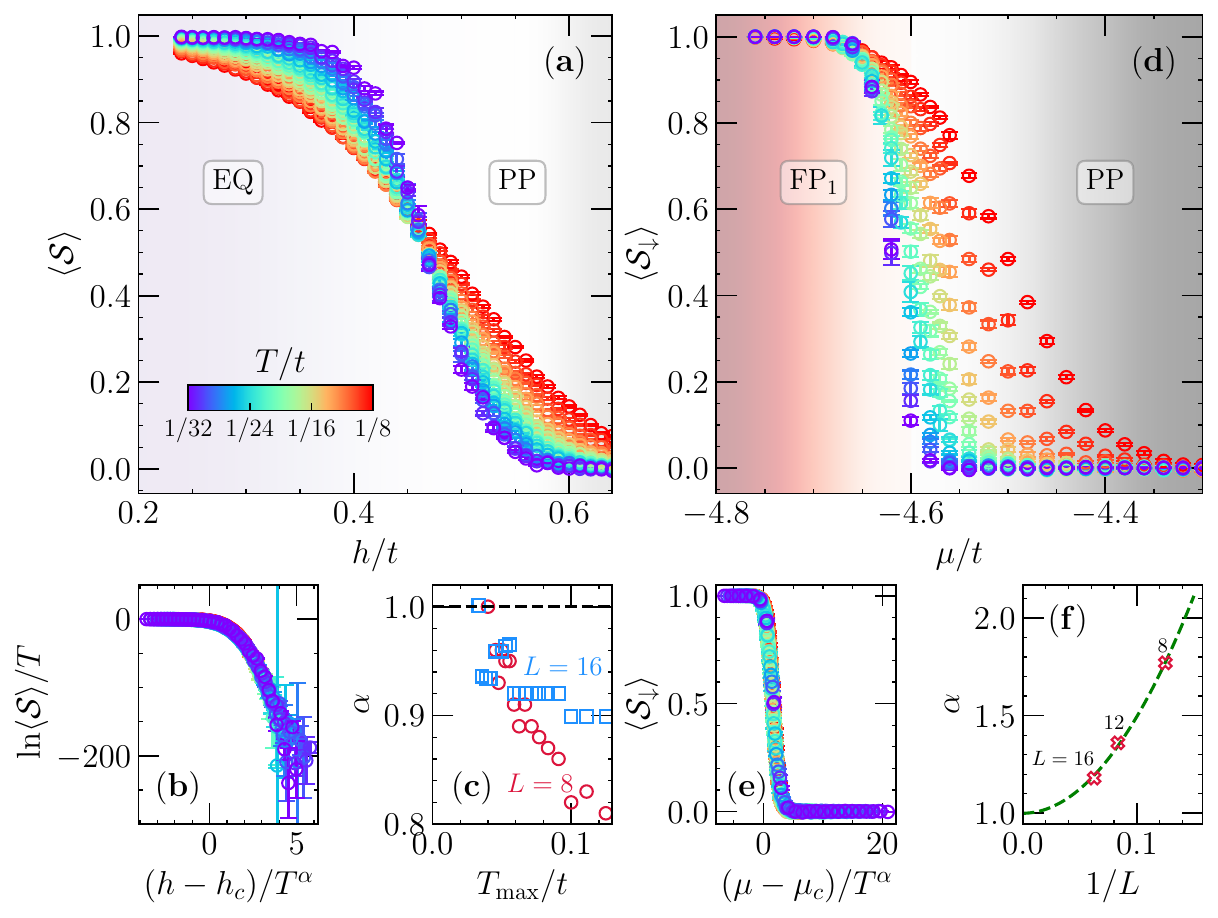}
\caption{Finite-$T$ criticality for the EQ-PP (a--c), $\mu/t=-5$, and the FP$_1$-PP (d--f), $h/t=2.5$, transitions, using $\langle {\cal S}\rangle$ and $\langle {\cal S}_\downarrow\rangle$, respectively. A $L=16$ lattice is used in (a) and (d). The best scaling collapse is shown in (b) and (e), $h_c=0.46t$ and $\mu_c=-4.66t$, respectively. The effects of the maximum temperature cutoff on the exponent $\alpha=(\nu z)^{-1}$ are shown in (c) for the EQ-PP transition (for two system sizes), and system size extrapolation in (f) for the FP$_1$-PP transition (with $T_{\rm max}/t = 1/8$); both low-$T_{\rm max}$ and large $L$ limits indicate $\alpha\to 1$.}
\label{fig:fig4}
\end{figure}

\paragraph{The two-dimensional case.---} Having established that phase boundaries to the PP phase can be tracked by the sign of different fermionic weights in small lattices, we now tackle a large 2D lattice with $L = 16 \ (L_x=L_y)$, as shown in Fig.~\ref{fig:fig3}. At $T/t=1/20$, the density [Fig.~\ref{fig:fig3}(a)] and imbalance [Fig.~\ref{fig:fig3}(b)] are similar to the results in Fig.~\ref{fig:fig1} but exhibit much smaller finite-size effects. Transitions to the trivial FP$_2$ and V phases are marked by ${\cal I}\to1$ ($\langle \hat n\rangle \to 1$) and ${\cal I}\to0$ ($\langle \hat n\rangle \to 0$), respectively. The average sign follows such behavior [Fig.~\ref{fig:fig3}(c)] in the same space of parameters, where we show line cuts in Figs.~\ref{fig:fig3}(g) and \ref{fig:fig3}(h) addressing $N$ finiteness. Indeed, the departure of $\langle {\cal S}\rangle$ from 1 marks the onset of FP$_1$ regimes from the trivial parent states, $V$ and FP$_2$, when approaching the thermodynamic limit -- we take $L_\tau/L$ approximately constant to preserve the aspect ratio of the $D+1$-dimensional lattice~\cite{Rieger1994, Mondaini2022, Mondaini2022-2}. The transition to the PP phase [Figs.~\ref{fig:fig3}(e) and \ref{fig:fig3}(f)], while typically less sharp due to finite-size and temperature effects, can be similarly inferred by having $\langle {\cal S}\rangle \to 0$ when transitioning from the EQ regime [Fig.~\ref{fig:fig3}(c)] or by $\langle {\cal S}_\downarrow\rangle \to 0$ when starting from the FP$_1$ phase [Fig.~\ref{fig:fig3}(d)].

\paragraph{Criticality.---} The importance of using the QMC weights to classify different physical regimes goes beyond a mere qualitative analysis the previous results suggest. As demonstrated in Ref.~\cite{Mondaini2022-2}, because the average sign $\langle{\cal S}\rangle$ is the ratio of partition functions ${\cal Z}_{W}/{\cal Z}_{|W|}$, any non-analyticity associated with critical behavior in ${\cal Z}_{W}$ is reflected in $\langle{\cal S}\rangle$ so long the auxiliary partition function ${\cal Z}_{|W|}$ is sufficiently analytic in that domain. Similar reasoning can be applied to the spin-resolved average sign~\cite{Mondaini2022-2} (see \cite{SM} for details). To show that this is indeed the case, we focus on the transitions to the PP phase at finite-but-low $T$'s. It is known that they exhibit free-fermion criticality, i.e., with dynamic and correlation critical exponents $z=2$ and $\nu=1/2$, respectively~\cite{Sachdev2011, Cheng2018, Cheng2018b, Luo2023, Luo2023b}. An interpretation for this type of criticality stems from the noninteracting composition of FFLO-like quantum liquids of paired and unpaired fermions that characterize the low-energy physics of the model in the strongly interacting regime~\cite{Kajala2011, Cheng2018}. In particular, the free energy for this type of transition, in the thermodynamic limit, reads ${\cal F} = T^{d/z+1}f\left(\frac{x-x_c}{T^{1/\nu z}}\right)$~\cite{Sachdev2011}, where $d$ is the physical dimension, $x=\mu$ or $h$ and $f$ is a scaling function~\cite{Zhou2010, Hazzard2011}. As a result, the above-mentioned relation of the average sign with the partition function yields that $\ln (\langle {\cal S}\rangle)/T \propto f\left(\frac{x-x_c}{T^{1/\nu z}}\right)$.

\begin{figure}[t!]\centering
\includegraphics[width=0.99\columnwidth]{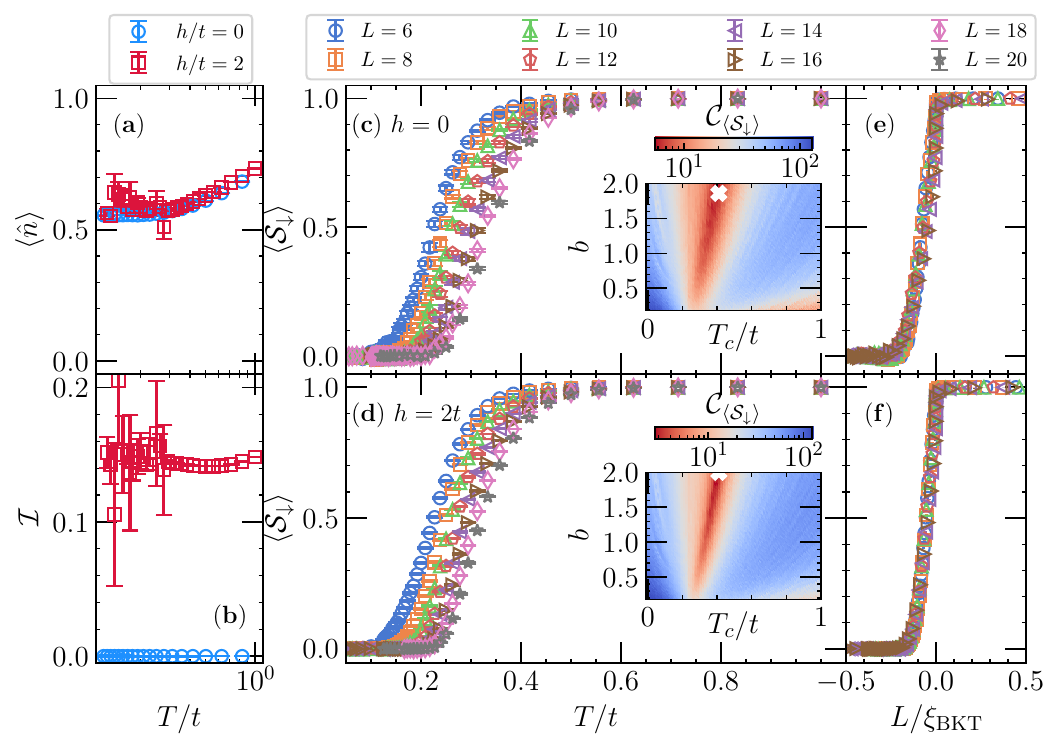}
\caption{The temperature dependence of the density (a) and imbalance (b) on a $L=20$, 2D lattice for $h/t=0$ and 2 with $\mu/t=-4$. Focusing on these field values, (c--f) give the BKT scaling analysis of $\langle {\cal S}_\downarrow\rangle$, where the $x$-axes in (e,f) are scaled by $L/\xi_{\rm BKT}$ ($-L/\xi_{\rm BKT}$) for $T>T_c$ ($T<T_c$). Insets give the cost function ${\cal C}_{\langle {\cal S}_\downarrow\rangle}$ of the scaling collapse in the space of parameters $(b,T_c)$ -- see \cite{SM} for definition. The white marker gives the best set of parameters used thus in (e) and (f).}
\label{fig:fig5}
\end{figure} 

Figure~\ref{fig:fig4} shows this scaling analysis using the average sign ($\downarrow$-spin average sign) for the EQ-PP (PP-FP$_1$) transition in a $L=16$ lattice. It highlights that at low temperatures [Fig.~\ref{fig:fig4}(c)] and approaching the thermodynamic limit [Fig.~\ref{fig:fig4}(f)], the best scaling collapse yields an exponent $\alpha \equiv (\nu z)^{-1}$ converging to 1, confirming this type of universality for such transitions in the model \eqref{eq:H}. 

We notice, however, that this does not establish superfluidity per se. Still, for either polarized or unpolarized systems, condensation of fermionic pairs entails spontaneous symmetry breaking of U(1) symmetry, which is only compatible, at finite temperatures for two-dimensional systems, with a Berezinskii-Kosterlitz-Thouless (BKT)-type transition~\cite{MerminWagner1966, Hohenberg1967}. Carrying over an analysis already used in the balanced case for the doped attractive Hubbard model~\cite{Mondaini2022-2}, we show that with a finite Zeeman-field, one can also obtain a scaling of QMC weights using a BKT-form. That is, using a correlation length $\xi_{\rm BKT}\propto \exp\left(b/\sqrt{T-T_c}\right)$, where $b$ is a non-universal constant and $T_c$ the critical temperature tracking the non-analyticity in the QMC weight (and, in turn, related to a phase transition to a quasi-superfluid regime), $\langle {\cal S}_\sigma\rangle$ exhibits scaling collapse, i.e., $\langle {\cal S}_\sigma\rangle = g(L/\xi_{\rm BKT})$.  

This is shown in Fig.~\ref{fig:fig5}, focusing on densities right above quarter-filling, contrasting zero and finite polarization, $h/t=0$ and 2. The scaling collapse of $\langle {\cal S}_\downarrow\rangle$ vs.~$L/\xi_{\rm BKT}$ with different system sizes [Fig.~\ref{fig:fig5}(e,f)] suggests a phase transition to a superfluid regime with a finite $T_c$. This is evidence that an FFLO phase is indeed robust in the 2D attractive Hubbard model, corroborating the $T=0$ results of Ref.~\cite{Vitali2022} and the finite-temperature ones of Ref.~\cite{Gukelberger2016}, yet employing the emergence of the sign problem for that.

\paragraph{Discussion.---} Using the critical properties of the average weights of the Monte Carlo sampling to identify phase transitions is becoming a recurring theme in QMC simulations~\cite{Mondaini2022-2, Mou2022}. A main caveat in this analysis is that the average sign depends on how interactions are decoupled or even on the basis on which one writes down the Hamiltonian, signifying that the partition function ${\cal Z}_{|W|}$ of the auxiliary system is not uniquely determined (see, however, \cite{SM} for a different HS transformation with similar critical properties). Thus, no guarantee exists that ${\cal Z}_{|W|}$ will be sufficiently analytic in the vicinity of the critical region, a condition for a scaling analysis to be meaningful. 

While this is a fundamental problem, it is plausible that finding a decoupling transformation (or basis) in which ${\cal Z}_{|W|}$ is acceptably controlled in the regime that ${\cal Z}_{W}$ exhibits critical behavior is a much simpler problem than finding a method where ${\cal Z}_{W} = {\cal Z}_{|W|}$ for the Hamiltonian under investigation. In other words, one can always make the SP much worse, ultimately masking criticality. Still, one does not need to solve it to infer the corresponding critical behavior, as we show here.

\begin{acknowledgments}
\paragraph{Acknowledgments.---} R.M.~gratefully acknowledges R.~T.~Scalettar for insightful discussions and a careful reading of the manuscript; R.M.~is also acknowledges support from the NSFC Grants No.~NSAF-U2230402, No.~12111530010, No.~12222401, and No.~11974039. 
T.-C.Y.~acknowledges support from the Science Foundation of Zhejiang Sci-Tech University Grant No.~23062182-Y.
Numerical simulations were performed in the Tianhe-2JK at the Beijing Computational Science Research Center and HSE HPC resources~\cite{Kostenetskiy2021}.
\end{acknowledgments}

\bibliography{refs}

\beginsupplement

\clearpage

\renewcommand{\theequation}{S\arabic{equation}}
\setcounter{equation}{0}

\onecolumngrid

\begin{center}

{\large \bf Supplementary Materials:
 \\ Two-dimensional polarized superfluids under the prism of the fermion sign problem }\\

\vspace{0.3cm}

\end{center}

\vspace{0.6cm}

\twocolumngrid
In these Supplementary Materials, we provide additional data to support the findings in the main text, including the study of different types of Hubbard-Stratonovich (HS) transformations, scaling of the sign of other weights, when possible, a description of the onset of trivial phases, and lastly an explanation of the finite-size effects that plague the regime of large polarization within the FP$_1$ phase.

\section{Scaling of the sign of other weights}
\begin{figure}[b]\centering
\includegraphics[width=0.99\columnwidth]{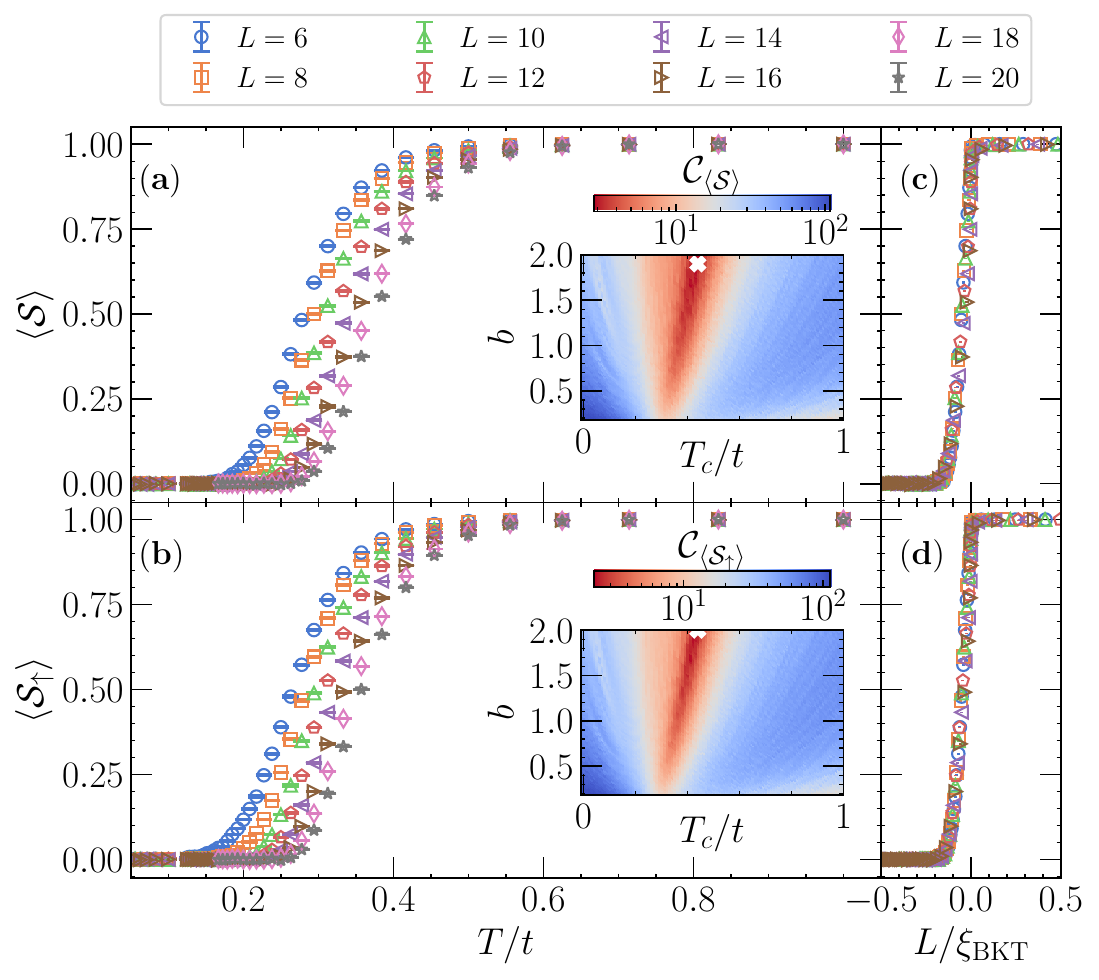}
\caption{Similar to Fig.~\tcr{5} in the main text, but now exhibiting a scaling analysis of the average sign $\langle {\cal S}\rangle$ and $\langle {\cal S}_\uparrow\rangle$ for a polarized regime with $h=2t$ and $\mu=-4t$. As before, the inset gives the cost function associated with the scaling collapse. It leads to compatible critical temperatures irrespective of the weight one is investigating.}
\label{fig:fig_s_sup_BKT}
\end{figure} 

In the main text, we argued that the transition to a superfluid can be signaled by the BKT-type scaling that the average sign of the weights must display. In particular, we focused on the sign of the $\downarrow$-spin component to directly compare the unpolarized ($h = 0$) and polarized ($h\neq 0$) regimes. Since the spin-resolved weights are precisely the same in the former case in a `charge decomposition' [or U(1) symmetric] scheme ~\cite{Hirsch1983,Hirsch1985} for whichever HS configurations, so are their signs. As a result, while one can similarly perform a scaling of $\langle {\cal S}_\uparrow\rangle$, the same cannot be done for the sign of the total weight since $\langle {\cal S}\rangle=1$ for all temperature and system sizes.

Specifically, that the spin-resolved average sign tracks the non-analyticity of phase transitions can also be easily seen by following a similar argument used for the total average sign. That is, starting from the definition of the average sign in terms of the ratio of partition functions,
\begin{equation}
    \langle {\cal S}\rangle = \frac{\sum_{\{x_{i\tau}\}} W(\{x_{i\tau}\})}{\sum_{\{x_{i\tau}\}}|W(\{x_{i\tau}\})|} = \frac{{\cal Z}_{W}}{{\cal Z}_{|W|}}\ ,
    \label{eq:sign_as_ratio_of_Zs}
\end{equation}
one can similarly obtain the average of a single fermionic flavor (suppose for simplicity 1 and 2) when they are separable as
\begin{align}
    \langle {\cal S}_1\rangle =&\frac{\sum_{\{x_{i\tau}\}} {\rm sgn}(W_1)\cdot |W_1W_2|}{\sum_{\{x_{i\tau}\}}|W_1W_2|} \times \frac{{\cal Z}_W}{{\cal Z}_W}\notag \\
    =& \langle {\cal S}\rangle \times \frac{\sum_{\{x_{i\tau}\}} {\rm sgn}(W_1)\cdot |W_1W_2|}{{\cal Z}_W}\ .
\end{align}
Here, in the first equality, we multiply by one for convenience; in the second equality, we use Eq.~\eqref{eq:sign_as_ratio_of_Zs}. As a result, one can rewrite the relation between original and reference partition functions as
\begin{equation}
    {\cal Z}_W = \frac{\langle {\cal S}\rangle}{\langle {\cal S}_1\rangle}\cdot {\cal Z}^\prime\ \text{where}\  {\cal Z}^\prime \equiv \sum_{\{x_{i\tau}\}} {\rm sgn}(W_1)\cdot |W_1W_2|\ .
\end{equation}
For the particular case of the BKT transition of an unpolarized superfluid, we notice that the only way in which one could track the associated non-analyticity in ${\cal Z}_W$ is via the flavor-resolved sign, since $\langle {\cal S}\rangle=1$, and thus ${\cal Z}_W = (1/\langle {\cal S}_1\rangle) \cdot {\cal Z}^\prime$.

This is not a restriction for the $h\neq 0$ regime, as no protection guarantees the individual weights to be equal. Following this, we display in Fig.~\ref{fig:fig_s_sup_BKT} the scaling of $\langle {\cal S}\rangle$ and $\langle {\cal S}_\uparrow\rangle$ (complementing the analysis of $\langle {\cal S}_\downarrow\rangle$ in Fig.~\tcr{5} in the main text) describing the BKT transition to a polarized superfluid for $h/t=2$ with $\mu/t=-4$, which results in densities close to quarter-filling [$\langle \hat n\rangle \gtrsim 0.5$] and imbalance ${\cal I} \simeq 0.15$ at low temperatures [see Fig.~\tcr{5}(a) and \tcr{5}(b)]. The results show that irrespective of the weight, the minima of the scaling collapse, resulting in a critical temperature $T_c$, is approximately the same.

In particular, the cost function to quantify the collapse we employ here and in the main text is given by ${\cal C}_y = \sum_j (|y_{j+1} - y_j|)/(\max\{y_j\} - \min\{y_j\})-1$, where $y_j$ are the values of either $\langle {\cal S}\rangle$ or $\langle {\cal S}_\sigma\rangle$, ordered according to their $L/\xi_{\rm BKT}$ ratio. Similar usage of this type of cost function for a BKT-transition has been applied in various contexts, including in phase transitions of infinite-temperature states~\cite{Suntajs2020,Aramthottil2021} or ground-state ones~\cite{Liang2023} in disordered systems.

\section{Different Hubbard-Stratonovich transformation}
As the `Discussion' section in the main text exposed, a main problem with the argument that the average sign of the weights tracks non-analytical behavior (and thus phase transitions) is that there is no unique form over which one can define $\langle {\cal S}\rangle$. That is, if changing basis or the Hubbard-Stratonovich transformation, one obtains a different reference partition function ${\cal Z}_{|W|}$, which can eventually mask any potential criticality in ${\cal Z}_{W}$, highlighting that the average sign cannot be interpreted as a physical quantity (or equivalently to say that it is not a gauge-invariant quantity)~\cite{Yan2023}. 

Despite that, to show that our results regarding criticality still hold, we employ another often-used HS transformation~\cite{Meng10}
\begin{align}
&e^{-\Delta \tau U (\hat n_{i\uparrow} + \hat n_{i\downarrow} - 1)^2/2} = \nonumber \\ &\sum_{x_{i\tau}=\pm1,\pm2} \gamma(x_{i\tau}) \prod_{\sigma} e^{{\rm i} \sqrt{\Delta\tau U/2}\eta(x_{i\tau})(\hat n_{i\sigma} -1/2)}+{\cal O}(\Delta\tau^4)\ ,\label{eq:SU2_HS}
\end{align}
which introduces a four-valued discrete field $x_{i\tau}=\pm1,\pm2$, with the real constants,
\begin{align}
    &\gamma(\pm1)=1+\sqrt{6}/3\ ;\  
    \eta(\pm1)=\pm\sqrt{2(3-\sqrt{6})} \notag\\
    &\gamma(\pm2)=1-\sqrt{6}/3\ ; \ 
    \eta(\pm2)=\pm\sqrt{2(3+\sqrt{6})}\ .
\end{align}
The transformation is not exact, but an error of [$\propto{\cal O}(\Delta\tau^4)$] is introduced, which is a small correction to the one that arises from the Trotter decomposition [${\cal O}(\Delta\tau^2)$]~\cite{Hirsch1985, dosSantos03}. Yet, to mitigate this extra systematic error, we reduce the imaginary time-discretization from $t\Delta\tau =0.10$ used in the U(1) HS transformation to $t\Delta\tau =0.05$ in this SU(2) case.

\begin{figure}[t]\centering
\includegraphics[width=0.99\columnwidth]{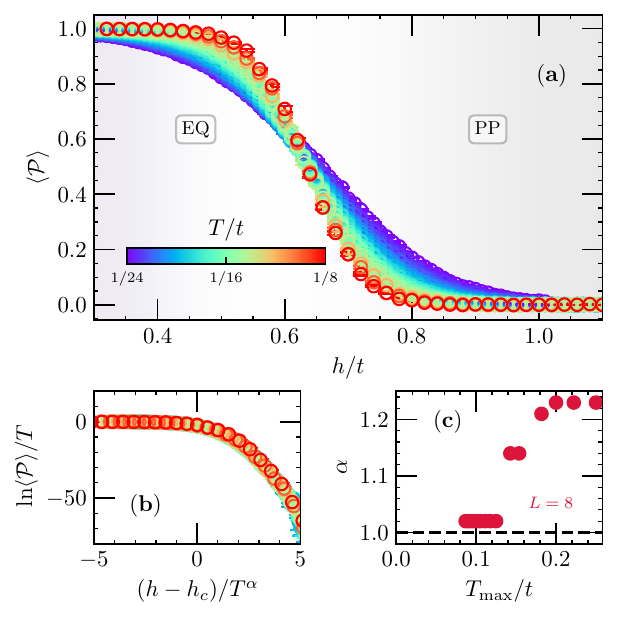}
\caption{Scaling analysis of the low-$T$ criticality of the EQ-PP transition using an SU(2) HS transformation, similar to Fig.~\tcr{4}(a--c) in the main text. The parameters are $L=8$, $\mu/t=-4$; critical field $h_c=0.54t$ in (b).}
\label{fig:fig_s_SU2_EQ_PP}
\end{figure}

\begin{figure}[t]\centering
\includegraphics[width=0.99\columnwidth]{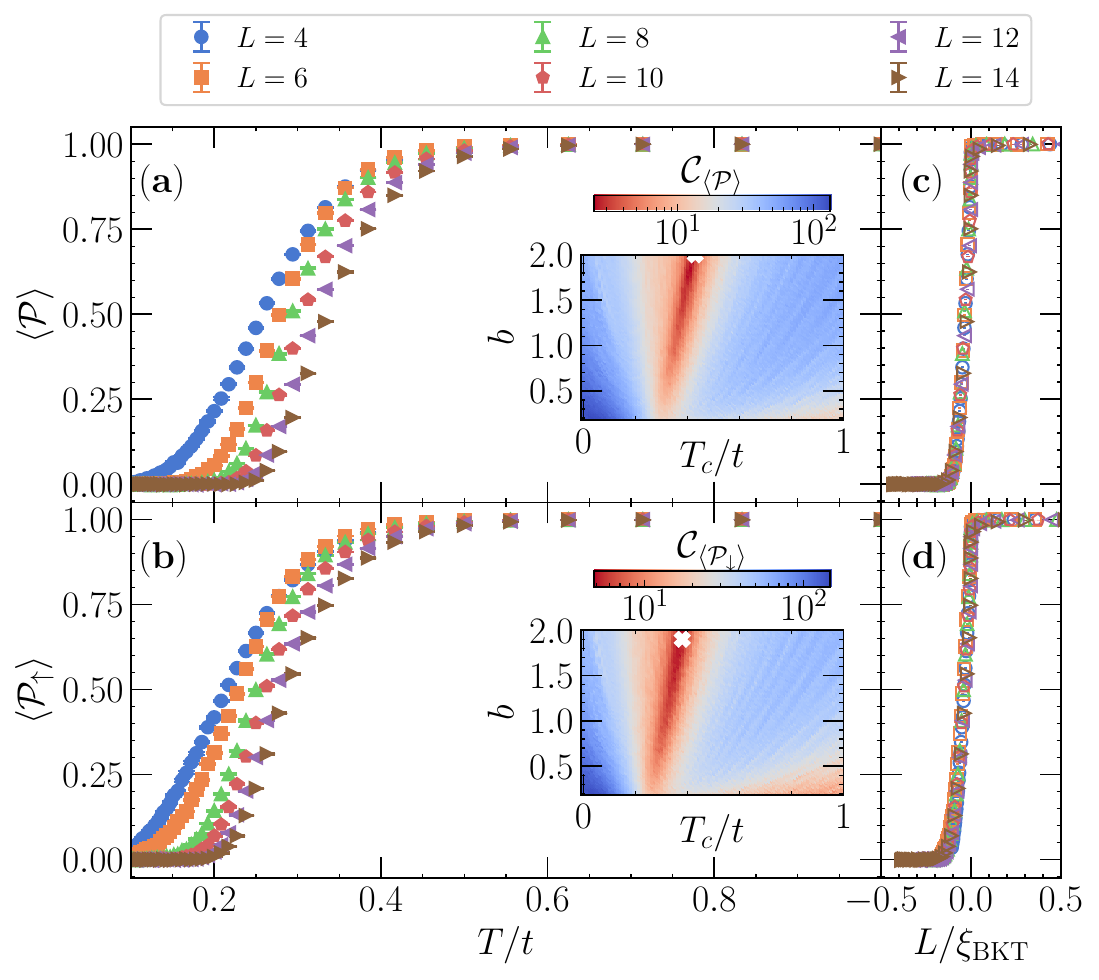}
\caption{BKT scaling characterizing the emergence of a quasi-long range superfluid at $h=2t$, $\mu=-4t$, here using an SU(2) HS transformation. Owing to the nature of the weights, which can be complex in principle, we plot the average value of the phases of the weights that converge to real values throughout the sampling (see text). As before, the quality of the scaling collapse is measured by the cost function in the corresponding insets.}
\label{fig:fig_s_SU2_BKT}
\end{figure}

We focus thus on two critical behaviors, the one at low-$T$ when crossing from EQ to PP phases [similar to Figs.~\tcr{4}(a--c) in the main text], and the BKT-type transition at finite Zeeman field that signals the onset of a quasi-long ranged polarized superfluid [as Figs. \tcr{5}(d) and \tcr{5}(f) in the main text]. The former is shown in Fig.~\ref{fig:fig_s_SU2_EQ_PP}, whereas the latter, in Fig.~\ref{fig:fig_s_SU2_BKT}. Here, we investigate the average phase of the weights,
\begin{equation}
    \langle {\cal P}\rangle =
    \frac{\sum_{\{x_{i\tau}\}} {\cal P}(W)\cdot |W(\{x_{i\tau}\})|}{\sum_{\{x_{i\tau}\}} |W(\{x_{i\tau}\})|}\ ,
\end{equation}
where ${\cal P}(W)$ is the (in principle) complex phase of the weight $W$ (i.e., $W = e^{{\rm i}\theta} |W| \equiv {\cal P}(W)\cdot|W|$). These turn out to be real throughout the sampling, and we proceed, as before, with the corresponding scaling analysis.

The criticality at low temperatures in the EQ-PP transition belongs to the free-fermion universality class. We proceed, as in the main text, with a scaling of $\ln (\langle {\cal P}\rangle)/T \propto f\left(\frac{h - h_c}{T^\alpha}\right)$, where $\alpha\equiv (\nu z)^{-1}$. Such universality class is characterized by the exponents $\nu=1/2$ and $z=2$ (i.e., $\alpha =1$)~\cite{Sachdev2011}; we observe that the scaling analysis, even for a small lattice with $L=8$, is sufficiently good to result in $\alpha \to 1$ when filtering it to consider lower largest temperatures $T_{\rm max}$, see Fig.~\ref{fig:fig_s_SU2_EQ_PP}(c), the regime where one expects the effects of criticality at finite $T$'s to be manifest.

Subsequently, the finite-$T$ BKT-type transition at finite Zeeman field (here with $h/t=2$) using the SU(2) HS transformation for either the total $\langle {\cal P}\rangle$ or partial $\langle {\cal P}_\sigma\rangle$ phases is analyzed in Figs.~\ref{fig:fig_s_SU2_BKT}(a) and \ref{fig:fig_s_SU2_BKT}(b), respectively. Again, scaling with the BKT correlation length $\xi_{\rm BKT}\propto \exp\left(b/\sqrt{T-T_c}\right)$ results in typical critical temperatures which are comparable to the ones obtained using a U(1) HS transformation in the main text. Here and before, note the relatively wide range in the cost function maps where the non-universal exponent $b$ results in sufficiently good scaling. Only the inclusion of substantially larger system sizes is likely to estimate the best set of parameters $(b, T_c)$ more accurately, and these are particularly more demanding in the SU(2) HS case owing to the complex matrices involved in the quantum Monte Carlo simulation.

\section{Transition to trivial phases}
Obtaining the loci for the onset of trivial phases (vacuum and fully polarized) follows directly by considering the $T=0$ energetics:
\begin{equation}
    E_0(N_\uparrow, N_\downarrow, \mu, h) = E_0(N_\uparrow, N_\downarrow) - \mu (N_\uparrow + N_\downarrow) - h(N_\uparrow - N_\downarrow)
    \label{eq:energetics}
\end{equation}
where $E_0(N_\uparrow, N_\downarrow)$ is the ground-state energy with a given number of spins-up $N_\uparrow$ and spin-down $N_{\downarrow}$. For example, the boundary between the vacuum and the EQ (equal densities) phase is made by comparing the corresponding energies, 0 and $E_0(1,1) - 2\mu$, respectively. The latter is the smallest possible filling satisfying the EQ constraint, which thus departs from the vacuum regime. It leads thus to $\mu = \frac{E_0(1,1)}{2}$, as annotated in Fig.~\tcr{1}(a) in the main text. Similarly, the boundary between the vacuum and FP$_1$ phases is obtained by comparing their energies according to Eq.~\eqref{eq:energetics}, 0, and $E_0(1,0)-h-\mu$, respectively. The latter is the smallest possible filling that leads to the fully polarized regime, leading to $\mu=-h+E_0(1,0)$. Here, the energy of a single particle is easily extracted; for a two-dimensional lattice with periodic boundary conditions, 
\begin{eqnarray}
E_0(1, 0)=\min\{-2t(\cos k_x+\cos k_y)\}=-4t.    
\end{eqnarray}
namely, $\mu = - h +4t$. Finally, the transition boundary between FP$_1$ and FP$_2$ phases can be similarly inferred (further employing a particle-hole transformation) to lead to $\mu = -h - E_0(1,0)$. Apart from those, we note that the remaining boundaries can be either estimated via finite-difference methods~\cite{Potapova2023} or via comparing all possible energy sectors of the different fillings of the Hamiltonian, as we have done to compile the boundaries for the $4\times 4$ lattice in Fig.~\tcr{1} in the main text. For the DMRG calculations used in Fig.~\tcr{2}, owing to a large number of possible fillings, we promoted a narrow search using indications for the density $\langle \hat n\rangle$ and polarization ${\cal I}$ extracted from the QMC calculations to extract the phase boundaries via similar energetics analysis as we show below.

\section{Transition to PP phase: DMRG vs.~QMC}
In the main text, Figure~\tcr{2}, we present a comparison of the DMRG results and the ones extracted from QMC in a $4\times 40$ ladder with open boundary conditions. We will here further explore how this benchmark is performed since the results are obtained with canonical and grand-canonical algorithms, respectively.

\begin{figure}[t]\centering
\includegraphics[width=0.99\columnwidth]{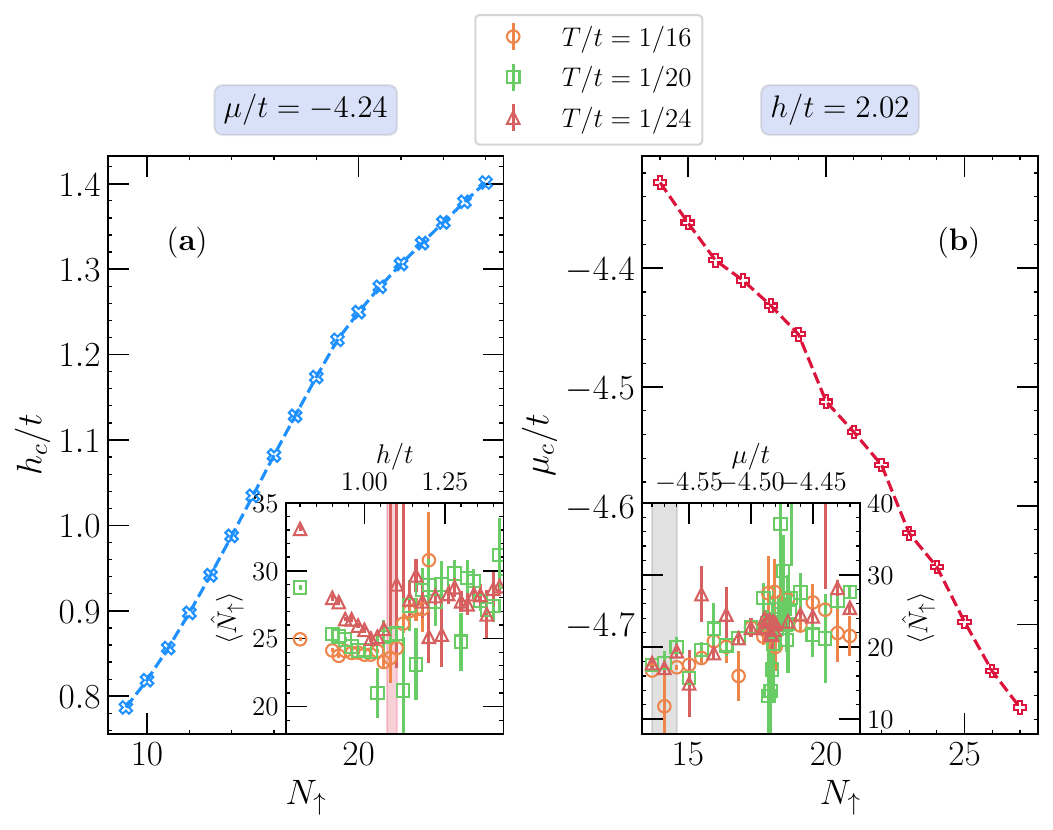}
\caption{(a) Critical Zeeman field as a function of the number of up spins $N_\uparrow$ when transitioning from an EQ to a PP phase for a fixed chemical potential $\mu/t=-4.24$. (b) Value of chemical potential that triggers an FP$_1$-PP transition for a fixed $h/t=2.02$ vs.~the number of up spins. The insets give the average density $\langle \hat N_\uparrow \rangle$ as a function of the driving parameters. Note the large error bars precisely because of the emergence of the sign problem. Shaded areas show an estimation of the transition location, given by $\langle {\cal S}\rangle\to0$ in the EQ-PP transition and when $\langle {\cal S}_\downarrow\rangle\to1$ in the FP$_1$-PP one.}
\label{fig:fig_sm_DMRG}
\end{figure} 
In DMRG calculations, we obtain the ground state for a given number of $N_\uparrow$ and $N_\downarrow$ spins, $E_0(N_\uparrow, N_\downarrow)$. The grand-canonical contribution thus reads, $E_0^{\rm x}=E_0(N_\uparrow, N_\downarrow) - \mu(N_\uparrow+N_\downarrow) - h(N_\uparrow-N_\downarrow)$; here $\rm x$ identifies the different phases studied in the main text. For the EQ-PP transition, we can compare the energy of a given balanced configuration with the one extracted with the smallest polarization $N_\uparrow\to N_\uparrow+1$. For example, $E_0^{\rm EQ} = E_0(N_\uparrow, N_\uparrow) - 2\mu N_\uparrow$ and $ E_0^{\rm PP (EQ+1)}=E_0(N_\uparrow+1, N_\uparrow)- \mu (2N_\uparrow + 1)-h$ which results in $h_c = E_0(N_\uparrow+1, N_\uparrow) - E_0(N_\uparrow, N_\uparrow) -\mu$ when these energies are made to be equal.

Figure~\ref{fig:fig_sm_DMRG} shows the corresponding critical Zeeman field when selecting the chemical potential to be $\mu/t=-4.24$, the value chosen in the QMC simulations in Fig.~\tcr{2}(a). In turn, the inset gives the average up spin occupancy $\langle \hat N_\uparrow\rangle$ using QMC at low temperatures. We notice that the prediction of the region where $\langle {\cal S}\rangle\to 0$, shaded regions, marks the onset of large fluctuations in $\langle \hat N_\uparrow\rangle$. Here, the corresponding values of $\langle\hat N_\uparrow\rangle$ fluctuate $\sim 19-25$. For $N_\uparrow = 19$, the energetic analysis of the critical field in the canonical simulation gives $h_c/t \simeq 1.21t$ [Fig.~\ref{fig:fig_sm_DMRG}(a)]. 

A similar analysis can be conducted in the FP$_1$-PP transition. Comparing the energy of a given filling in the (non-saturated) full polarized regime, $E_0^{\rm FP_{1}} = E_0(N_\uparrow, 0)-\mu N_\uparrow-hN_\uparrow$, with the energy of a minimal configuration that renders the PP regime, $E_0^{\rm PP(FP_1+1)} = E_0(N_\uparrow, 1) - \mu(N_\uparrow +1)-h(N_\uparrow-1)$, results in $\mu_c = E_0(N_\uparrow, 1) - E_0(N_\uparrow, 0) + h$. As before, the region in which $\langle {\cal S}_\downarrow\rangle$ departs from 1 in the QMC simulations can be roughly pointed to $\mu/t \in [-4.58, -4.56]$. In this region, $\langle \hat N_\uparrow\rangle$ yields values up to 21 [see inset in Fig.~\ref{fig:fig_sm_DMRG}(b)]; the energetic analysis for this filling gives thus $\mu_c/t = -4.54$ when using the QMC based Zeeman fields of $h/t=2.02$. 

These critical boundaries should be taken as an approximate location of the critical boundary since, unlike the ED results for the $4\times 4$ lattice, we cannot directly compare all the filling sectors, determining which one has the lowest grand-canonical energy for this system size. Nonetheless, they provide sufficient evidence of the correlation of the partially polarized regime being tracked by the average sign of corresponding weights in the QMC sampling.

\section{Comparison of small lattices: Finite-size effects}
\begin{figure}[b!]\centering
\includegraphics[width=0.9\columnwidth]{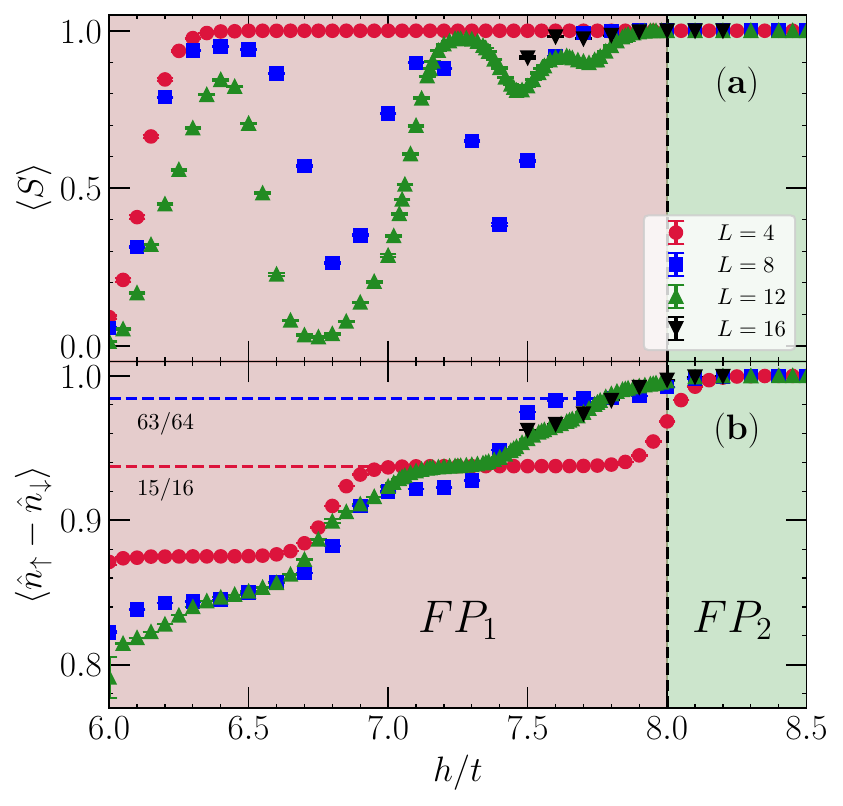}
\caption{The average sign $\langle {\cal S}\rangle$ (a) and the imbalance ${\cal I}$ across the FP$_1$-FP$_2$ transition with growing $h$ at $\mu/t = -4$ and $T/t = 1/20$.}
\label{fig:fig_fse}
\end{figure} 
In Figs.~\tcr{3}(c,d,g) of the main text, large fluctuations of the average signs $\langle {\cal S}\rangle$ and $\langle {\cal S}_\downarrow\rangle$ with growing Zeeman field can be observed, in particular close to the boundary to the trivial fully polarized phase FP$_2$. These readily follow the closed-shell fillings for the corresponding system sizes, i.e., a set of commensurate fillings at low temperatures show that $\langle {\cal S}\rangle\to 1$, while the average sign drops dramatically within the FP$_1$ phase once one transitions from one commensurate filling to another~\cite{Mondaini2012}. Remarkably, for the smaller lattices, the average sign starts converging towards one once the corresponding spin-$\uparrow$ density reaches values $(N-1)/N$, i.e., in regimes in which one has a single-hole embedded in a `sea' of $\uparrow$-polarized fermions. This is particularly clear for the $8\times 8$ lattice, with filling $\langle \hat n\rangle = \langle \hat n_\uparrow\rangle = 63/64$, see Fig.~\ref{fig:fig_fse}. Via a particle-hole transformation, this is equivalent to a one-particle situation for which one could not obtain a fermion sign problem. As a result, marking the onset of the FP$_2$ phase by when $\langle {\cal S}\rangle\to 1$ can be significantly compromised in a small lattice since such `one-particle' fillings can be extended in $h$ but quickly shrink once larger lattice sizes are considered.

\end{document}